\documentstyle[11pt,twoside,paspconf,epsfig]{article}

\begin{document}

\title{The Full-sky Astrometric Mapping Explorer -- Astrometry for the New
Millennium}

\author{
Scott D.\ Horner\altaffilmark{1},
Marvin E.\ Germain\altaffilmark{2},
Thomas P.\ Greene\altaffilmark{3},
Fred H.\ Harris\altaffilmark{2},
Mark S.\ Johnson\altaffilmark{4},
Kenneth J.\ Johnston\altaffilmark{1},
David G.\ Monet\altaffilmark{2},
Marc A.\ Murison\altaffilmark{1},
James D.\ Phillips\altaffilmark{5},
Robert D.\ Reasenberg\altaffilmark{5},
P.\ Kenneth Seidelmann\altaffilmark{1},
Sean E.\ Urban\altaffilmark{1},
and
Richard H.\ Vassar\altaffilmark{6}
}

\footnotesep 1ex

\altaffiltext{1}{U.S.\ Naval Observatory, 3450 Massachusetts Ave.\ NW,
Washington, DC 20392-5420}
\altaffiltext{2}{U.S.\ Naval Observatory, PO Box 1149, Flagstaff, AZ
86002-1149}
\altaffiltext{3}{NASA Ames Research Center, M.S.\ 245-6, Moffett Field, CA
94035-1000}
\altaffiltext{4}{Naval Research Laboratory, Code 8100, 4555 Overlook Ave.\ SW,
Washington, DC  20375-5000}
\altaffiltext{5}{Smithsonian Astrophysical Observatory, 60 Garden St.,
Cambridge, MA 02138}
\altaffiltext{6}{Lockheed Martin Missiles and Space Advanced Technology
Center, O/L923 B/251, 3251 Hanover St., Palo Alto, CA 94304-1191}

\begin{abstract}

FAME is designed to perform an all--sky, astrometric survey with
unprecedented accuracy.  It will create a rigid astrometric
catalog of 4 $\times 10^7$ stars with $5 < m_V < 15$.  For bright
stars, $5 < m_V < 9$, FAME will determine positions and parallaxes
accurate to $< 50\ \mu as$, with proper motion errors $< 50\ \mu
as/year$.  For fainter stars, $9 < m_V < 15$, FAME will determine
positions and parallaxes accurate to $< 500\ \mu as$, with proper
motion errors $< 500\ \mu as/year$.  It will also collect
photometric data on these 4 $\times 10^7$ stars in four Sloan DSS
colors.

\end{abstract}

\section{Introduction}

NASA selected the Full-sky Astrometric Mapping Explorer (FAME) to
be one of five MIDEX missions funded for a concept study.  The
Phase A Concept Study Report was submitted to NASA on 18 June
1999. In September 1999, NASA will select two of these five
missions for flight as MIDEX-3 (CY2003 launch) and MIDEX-4 (CY2004
launch) in its Explorer program.

While not an interferometer, FAME is relevant to interferometry
because of the way it complements interferometric projects, in
particular the Space Interferometry Mission (SIM, cf.\ Shao 1999).
The proper motion data from FAME, combined with {\it Hipparcos}
(cf.\ Kovalevsky 1998) and other data will be ideal for use to
select SIM's astrometric reference grid stars.  FAME will also
identify stars with nonlinear proper motions as planetary system
candidates for further study by SIM, Terrestrial Planet Finder,
and future ground based interferometers.  The fundamental
astrometric data provided at relatively low cost by FAME will help
optimize the scientific return from these future projects.  This
is in addition to the considerable, direct, scientific return from
FAME.  FAME will redefine the extragalactic distance scale and
provide a large, rich database of information on stellar
properties that will enable numerous science investigations in
stellar structure and evolution, the dynamics of the Milky Way,
and stellar companions including brown dwarfs and giant planets.

\section{Science Goals}

FAME will measure the positions, parallaxes, and proper motions of
4 $\times 10^7$ stars with $5 < m_V < 15$.  The positional
accuracy will be the finest yet achieved.  For $m_V = 9$ stars,
the positional and parallax accuracies will be better than $ 50\
\mu as$, and the proper motion accuracies will be better than $50\
\mu as/year$.  At $m_V = 15$, these accuracies will be degraded by
only an order of magnitude.  FAME will also obtain photometric
data with millimagnitude accuracies on these 4 $\times 10^7$
stars, observing them in four of the five Sloan Digital Sky Survey
bands (g$^\prime$, r$^\prime$, i$^\prime$, z$^\prime$).

\begin{figure}[t]
\plotone{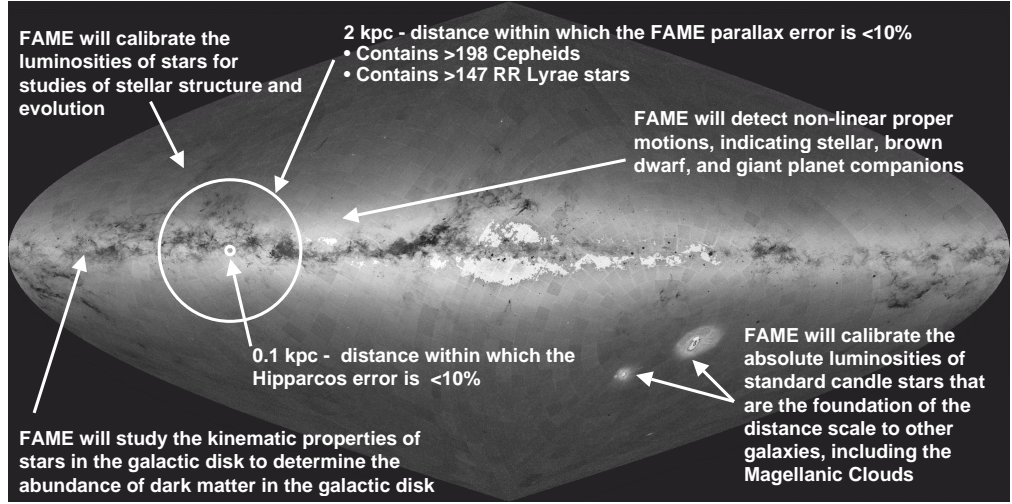}
\caption{The principal science goals of
FAME. It will not only improve on the accuracies of star positions
determined by Hipparcos but also expand the volume of space for
which accurate positions are known by a factor of 8000.}
\label{galaxy}
\end{figure}

Not only will the FAME data provide a rigid, accurate, optical,
astrometric grid, but it will also produce an extensive database of stellar
properties that will enable research in areas across the NASA themes.
Figure~\ref{galaxy} summarizes the principal scientific objectives of FAME.

\subsection{Extragalactic distance scale}
\nobreak

\begin{figure}[t]
\plotone{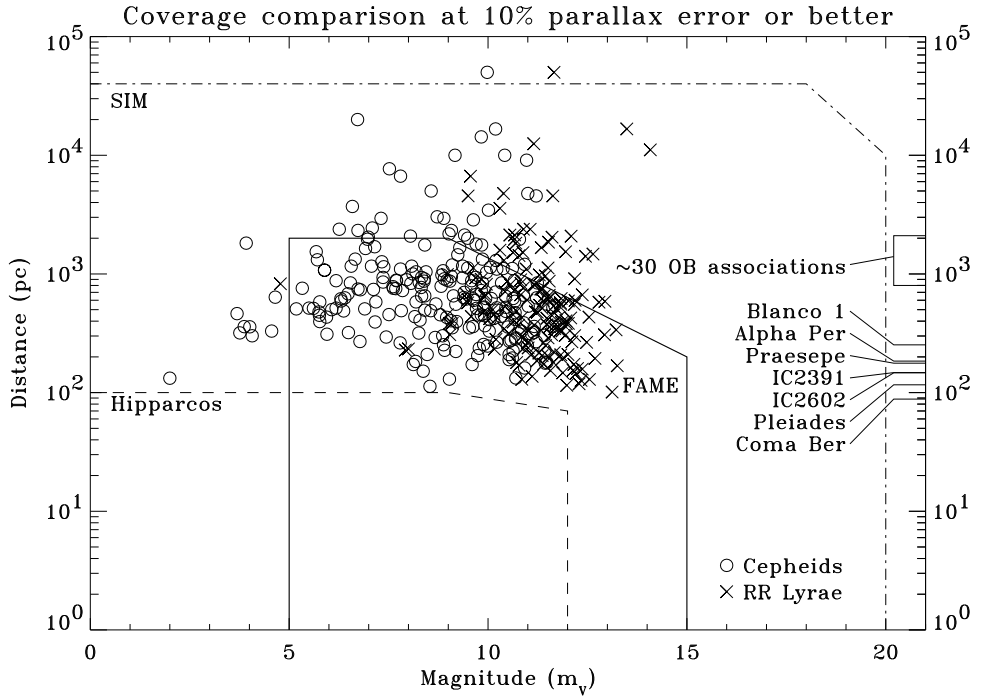}
\caption{Comparison of the astrometric
capabilities of FAME to SIM and Hipparcos.  The lines indicate
where achieved accuracy will be 10\% parallax error or better.
Known Cepheids are indicated by circles and RR Lyraes with
`$\times$'s. Distances to clusters and OB associations are
indicated at the right of the figure.}
\label{cepheids}
\end{figure}

FAME will calibrate the absolute luminosities of ``standard
candle'' stars such as Cepheids and RR Lyrae stars.
Figure~\ref{cepheids} shows the coverage of the Hipparcos, SIM,
and FAME missions within which distances are or will be accurate
to better than 10\% error. Also plotted in this figure are the
known Cepheids and RR Lyrae stars. FAME has been optimized to
obtain accurate distances to a large sample of Cepheids to
determine the zero--point of the period--luminosity relation,
calibrating the extragalactic distance scale.

\subsection{Stellar Companions \& Exoplanets}
\nobreak

FAME will detect low--mass companions by measuring nonlinear proper motions of
catalog stars.  This will provide a definitive determination of the frequency
of solar--type stars orbited by brown dwarf companions in the mass range of 10
to 80 $M_{jup}$ with orbital periods as long as about twice the duration of
the FAME mission.  This will include an exploration of the transition region
between giant planets and brown dwarfs, which appears to be in the range 10 to
30 $M_{jup}$.

\subsection{Absolute Luminosities}
\nobreak

By determining their parallaxes, FAME will calibrate the absolute luminosities
of solar--neighborhood stars, including Population I and Population II stars,
thus enabling diverse studies of stellar evolution and stellar structure.
In the case of Population II subdwarfs, this will allow the determination of
the distances and ages of galactic and extragalactic globular clusters with
unprecedented accuracy.

The FAME database will also include four--color photometry for
stars in the solar neighborhood.  This unparalleled wealth of
knowledge of fundamental stellar properties will revolutionize our
understanding of stars in our quadrant of the Milky Way.

\subsection{Dynamics}
\nobreak

By providing the proper motions and parallaxes of 4 $\times 10^7$
stars, FAME will enable studies of the kinematic properties of
solar neighborhood stars.  In particular, we can assess the
abundance and distribution of dark matter in the galactic disk
with much greater sensitivity and completeness than previously
possible.

We can also measure the proper motions and distances for individual stars in
star forming regions for determination of the ages and kinematics of those
regions.

\section{Mission Design}

FAME evolved from the highly successful {\it Hipparcos} design.
As with {\it Hipparcos,} FAME uses two widely separated fields of
view that are combined on a single focal plane to control the
growth of random errors in the relative separations of stars over
large angles.  Unlike {\it Hipparcos,} however, FAME will have a
large array of CCDs that will not only improve the signal to noise
of the observations but will also enable the observation of many
stars simultaneously.

\begin{figure}[t]
\plotone{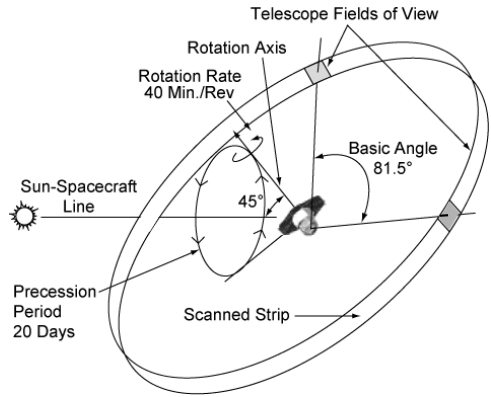} \caption{The rotation axis of the FAME
spacecraft is pointed 45$^{\circ}$ from the Sun and precesses
around the Sun with a 20 day period.  The FAME spacecraft rotates
with a 40 minute period.  The two fields of view are normal to the
rotation axis and are separated by a 81.5$^{\circ}$ basic angle.}
\label{scan}
\end{figure}

FAME will use the solar radiation pressure on the Sun shield/solar
array panels to smoothly precess its spin axis about the Sun,
maintaining a separation of about 45$^{\circ}$ between the spin
axis and the Sun direction. This will allow the reduction or
elimination of thruster firings for precession, which results in a
reduction of systematic errors introduced by breaks in the smooth
motion of the spacecraft. Figure~\ref{scan} illustrates the FAME
observing concept.

FAME will be launched by a Delta II 7425 to a geosynchronous
transfer orbit, then boosted into a geosynchronous orbit by an
apogee kick motor.  This orbit allows for 24-hour communication
with the spacecraft, reduces the thermal impact of the Earth on
the instrument, and reduces occultations and eclipses.

\section{Instrument}

The FAME instrument is designed, assembled, aligned, and tested by
Lockheed Martin Missiles and Space Advanced Technology Center at
its Palo Alto facility.  The instrument has a compound mirror,
consisting of two 0.6 m $\times$ 0.25 m flats mounted at a fixed
``basic angle'' of 81.5$^{\circ}$. The compound mirror reflects
the light from the two fields of view into a common optical train,
consisting of three powered surfaces and a set of five flats to
fold the 15 m focal length of the optical system into the
available volume. The resulting flat focal plane combines the
images from the two fields of view onto an array of twenty--four
2048 $\times$ 4096 backside illuminated CCDs.

The CCDs are clocked out in time delayed integration mode to match
the rotation rate of the spacecraft.  The instrument autonomously
compensates for variations in the spacecraft spin rate by
analyzing bright star data on--board to calculate the spin rate
and then adjusting the TDI rate accordingly.

\section{Spacecraft Bus}

The spacecraft bus is designed, assembled, integrated with the
instrument, and tested by the Naval Research Laboratory (NRL). The
bus will place the instrument in the proper orbit, provide a
long-term stable platform for the instrument, and collect, buffer,
and transmit the science data to the ground station.

To provide a stable platform for the instrument, all actively moving
components were eliminated.  The spacecraft bus thermal design and operation
modes are such that constant power and temperatures are maintained to
eliminate structural expansion or contraction.  Passive damping is employed to
maintain a low level of jitter.

Many of the subsystems used in FAME have flight heritage from {\it
Clementine.} The NRL Naval Center for Space Technology has built and launched
over 87 satellites since 1960, thus NRL's FAME team has extensive experience
with rapid spacecraft development.

\begin{figure}[t]
\plotone{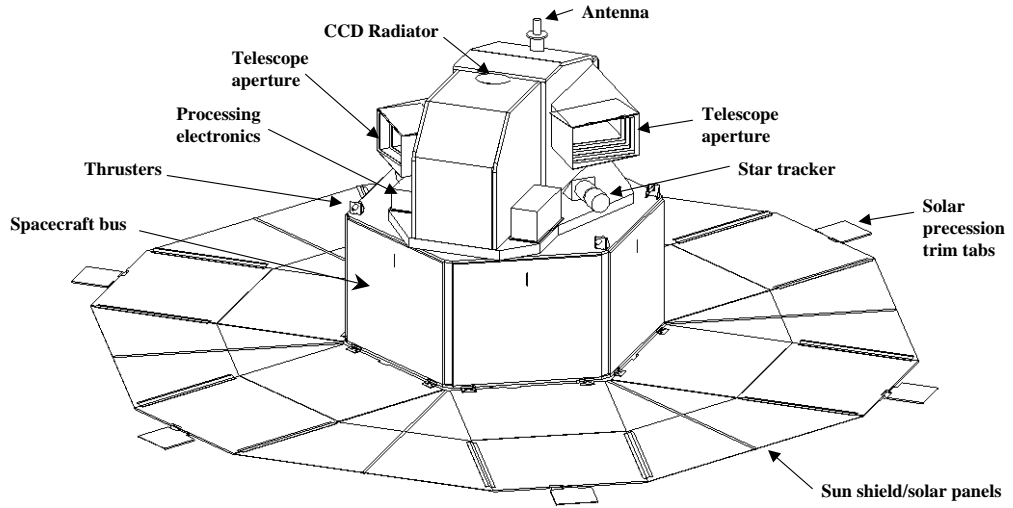}
\caption{The FAME spacecraft after solar shield deployment}
\label{sc}
\end{figure}

Figure~\ref{sc} shows the FAME spacecraft in its operational configuration.
The solar shield/solar array panels deploy and provide power to the
spacecraft, optically and thermally shield the instrument from the Sun, and
act as a solar sail to smoothly precess the spacecraft.  Motorized trim tabs
mounted on the edge of the solar shield can adjust the rate of precession for
optimal performance.

\section{Summary}

In addition to its primary science objectives, FAME is an extremely valuable
complement to the SIM mission.  FAME, an all--sky survey, naturally
complements SIM, a pointed mission.  SIM will obtain very high precision
observations of a limited number of objects ($<10,000$), whereas FAME will
obtain high precision observations of a large number of objects, providing the
statistical samples required for many studies (cf.\ Spergel 1999).

FAME provides a database of faint, astrometrically stable stars
for use in selecting the SIM grid stars.  This would be extremely
valuable in reducing the amount of time SIM will spend observing
the grid stars.  It will also identify targets with nonlinear
proper motions for observation with SIM to search for low--mass
companions and exoplanets.  SIM observing time will be expensive,
both in terms of the overall cost of the SIM mission and the
limited time available for the large number of key projects.  The
FAME database will reduce the amount of time required to observe
grid stars, and identify interesting targets for exoplanet
searches.

\acknowledgments

Funding for the FAME proposal development and MIDEX Phase A Concept Study
were provided by the U.S.\ Naval Observatory, NASA Explorers Program, and
Lockheed Martin Missiles and Space.

FAME is a joint development effort of the U.S. Naval Observatory, Lockheed
Martin Missiles and Space Advanced Technology Center, Naval Research
Laboratory, and Smithsonian Astrophysical Observatory.


\begin{references}

\reference Kovalevsky, J. 1998, \araa, 36, 99

\reference Shao, M. 1999, in ASP Conf.\ Ser., Working on the Fringe, eds.\
S.C.\ Unwin \& R.V.\ Stachnik (San Francisco: ASP), {\it these proceedings}

\reference Spergel, D. 1999, in ASP Conf.\ Ser., Working on the Fringe, eds.\
S.C.\ Unwin \& R.V.\ Stachnik (San Francisco: ASP), {\it these proceedings}

\end{references}
\end{document}